# The mystery of relationship of mechanics and field in the many-body quantum world

Michal Svrček

**Abstract**

We have revealed three fatal errors incurred from a blind transferring of quantum field methods into the quantum mechanics. This had tragic consequences because it produced crippled model Hamiltonians, unfortunately considered sufficient for a description of solids including superconductors. From there, of course, Fröhlich derived wrong effective Hamiltonian, from which incorrect BCS theory arose.
  1) Mechanical and field patterns cannot be mixed. Instead of field methods applied to the mechanical Born-Oppenheimer approximation we have entirely to avoid it and construct an independent and standalone field pattern. This leads to a new form of the Bohr's complementarity on the level of composite systems.
  2) We have correctly to deal with the center of gravity, which is under the field pattern "materialized" in the form of new quasipartiles - rotons and translons. This leads to a new type of relativity of internal and external degrees of freedom and one-particle way of bypassing degeneracies (gap formation).
  3) The possible symmetry cannot be apriori loaded but has to be aposteriori obtained as a solution of field equations, formulated in a general form without translational or any other symmetry. This leads to an utterly revised view of symmetry breaking in non-adiabatic systems, namely Jahn-Teller effect and superconductivity. These two phenomena are synonyms and share a unique symmetry breaking.

## 1. Introduction

Classical physics clearly distinguishes what is mechanics and what is field. We know classical mechanics formulated by Newton's laws, and the theory of the electromagnetic field determined by Maxwell's equations. In the quantum world, these two theories are reflected in two independent disciplines - quantum mechanics and quantum field theory.

When applied to many-body systems, quantum mechanics, however, unlike classical mechanics does not use purely mechanical means for formulating and solving equations, but takes over field methods from the quantum field theory. Thus, in full analogy with the electron-photon interaction, the concept of electron-phonon interaction was introduced in solid state physics, but in quantum chemistry field methods are usually used only on the electronic level [14-16].

In solid state physics, perhaps the most famous is the model field Hamiltonian

$$H = \sum_{\mathbf{k},\sigma} \varepsilon_{\mathbf{k}} a^+_{\mathbf{k},\sigma} a_{\mathbf{k},\sigma} + \sum_{\mathbf{q}} \hbar\omega_{\mathbf{q}} \left( b^+_{\mathbf{q}} b_{\mathbf{q}} + \tfrac{1}{2} \right) + \sum_{\mathbf{k},\mathbf{q},\sigma} u^{\mathbf{q}} \left( b_{\mathbf{q}} + b^+_{-\mathbf{q}} \right) a^+_{\mathbf{k+q},\sigma} a_{\mathbf{k},\sigma} \qquad (1)$$

In this paper, we focus on the analysis of this expression: how it was derived, and whether it is not only an incomplete expression of far more general and yet unnoticed field mechanisms within the quantum mechanics. We have a very good reason to do it. From the Hamiltonian (1) Fröhlich Hamiltonian [22] was derived and from it subsequently the BCS theory of superconductivity [23]. This theory is able to predict the superconducting materials approximately only upto 30°K. After the discovery of high-temperature superconductivity, hundreds of theories have appeared with dozens of exotic mechanisms, replacing the original electron-phonon one in the BCS theory. In several works also the Fröhlich Hamiltonian was reexamined, the most detailed in [28]. It turns out that the final form of the Fröhlich Hamiltonian is ambiguous, and offers a range of solutions, where the effective electron-electron Hamiltonian can be in different forms partially attractive and repulsive as the original Fröhlich's result, or even completely attractive. This finding, however, does not affect the concept on which the BCS theory is built.

The heart of the matter, namely the field Hamiltonian (1) has never been verified, if it is correct. It is considered as a model one, and therefore not accurate. Is its accuracy sufficient for capturing the essence of superconductivity? There are two known historical facts about Fröhlich and Bardeen. When Fröhlich fully realized the implications of his research that from his Hamiltonian the BCS theory was derived, according to his "common sense" he disagreed with the BCS theory and argued that superconductivity must be explained on the basis of one-particle theory. But it was too late. He himself was unsuccessful with the derivation of one-particle theory, so eventually the BCS theory survived. And again, when Bardeen saw the consequences of BCS theory after the discovery of Josephson effect [27], according to his "common sense" claimed that Cooper pairs cannot tunnel through the insulating layer. It was late again. There was no alternative of the BCS theory, and the admission of one-particle tunneling would lead to its denial. So finally Anderson's opinion won recognition.

Scientific knowledge is deeply anthropomorphic. Were not finally Fröhlich and Bardeen with their "common sense" right? If so, there must be something wrong in the implementation of the model field Hamiltonian (1). To check it, we have to make three generalizations: 1) to avoid completely the Born-Oppenheimer (B-O) approximation, 2) to analyze the centre-of-mass (COM) problem that was never considered in the field methods, and 3) to go beyond the restrictions resulting from the translational symmetry.

1) The Hamiltonian (1) was derived from the B-O approximation [3], and therefore is not able to describe non-adiabatic phenomena. Superconductivity is a clearly non-adiabatic phenomenon. Therefore this Hamiltonian has to be rederived on the non-B-O level, i.e. without using the B-O approximation.

2) Field methods, as implemented in solids, totally ignore the COM problem. Generalized field equations in the adiabatic limit must precisely agree with the next correction above the B-O approximation, i.e. the Born-Huang [4] ansatz. As it will be shown, the model

Hamiltonian in the form (1) does not meet this requirement at all, is not even able to explain adiabatic phenomena within the first correction of the B-O approximation. How, then, can it explain non-adiabatic phenomena? We address the problem here, how the quantum field should properly reconcile itself to the COM problem.

3) Crystals with translational symmetry, however, represent only a subset of condensed matter - molecules and solids. If looking for a general field archetype, it is necessary to generalize all derivations for cases without translational symmetry. This generalization is essential also in cases where translational symmetry exists. It is therefore necessary to work consistently with equations without translational symmetry, and to apply the translational symmetry no sooner than in the final expressions. Even though we are working with systems with translational symmetry, until the last moment we do not know what kind of symmetry it specifically will be. This is very important for a proper understanding of symmetry breaking, which occurs in the Jahn-Teller effect [6] and superconductivity.

## 2. Field Hamiltonian in the B-O approximation

Let us start with a brief recap of the B-O approximation and how the Hamiltonian (1) was derived from it.

Schrödinger equation for a general system of nuclei and electrons can be written as

$$H = T_N(\vec{R}) + T_e(\vec{r}) + E_{eN}(\vec{r},\vec{R}) + E_{NN}(\vec{R}) + E_{ee}(\vec{r}) \qquad (2)$$

where $T$ denotes the kinetic energies and $E$ the potential energies. Coordinates $\vec{R}$ and $\vec{r}$ and indices $e$ and $N$ refer to electrons and nuclei.

It is analytically impossible to solve this equation exactly. If the term $E_{eN}$ did not exist, equation (2) would lead to the separation of electron and nuclear motions, and the wave function $\Psi(\vec{r},\vec{R})$ of the system would be simply expressed as the product of the electron subsystem $\psi(\vec{r})$ and the nuclear subsystem $\chi(\vec{R})$.

B-O approximation [3] thus provides a simplified solution, based on the assumption that the masses of the electrons are much smaller than the masses of nuclei. It is based on the expansion proportional to the ratio $m/M$, where the nuclei are almost fixed with respect to the movement of electrons; only electrons parametrically reflect the movement of nuclei. The wave function is thus expressed as

$$\Psi(\vec{r},\vec{R}) = \psi(\vec{r},\vec{R}) \chi(\vec{R}) \qquad (3)$$

So the B-O approximation assumes that the separation of electrons and nuclei can be performed, albeit only in an approximative way with the parametric dependency $\psi(\vec{r},\vec{R})$ on $\vec{R}$. Therefore it is possible to create potential curves by the calculation of the $\psi(\vec{r},\vec{R})$ depending on the actual position of nuclei $\vec{R}$. If we denote the electronic part of the Hamiltonian as $H_e$

$$H_e = T_e(\vec{r}) + E_{eN}(\vec{r},\vec{R}) + E_{ee}(\vec{r}) \qquad (4)$$

we obtain for the electron states the equation

$$H_e(\vec{R})|\psi(\vec{r},\vec{R})\rangle = E_e(\vec{R})|\psi(\vec{r},\vec{R})\rangle \qquad (5)$$

and for the nuclear states the equation

$$H_N|\chi(\vec{R})\rangle = E|\chi(\vec{R})\rangle \qquad (6)$$

where in the first approach neglecting off-diagonal adiabatic terms we get

$$H_N = T_N(\vec{R}) + E_{NN}(\vec{R}) + E_e(\vec{R}) + \langle\psi(\vec{R})|T_N|\psi(\vec{R})\rangle \qquad (7)$$

The first three terms in (7) represent the B-O approximation, the last term the diagonal adiabatic correction, called the Born-Huang [4] or Born-Handy ansatz [5].

How is the equation (1) tied in with B-O approximation? Electron and electron-phonon terms come from the second quantization of equation (4) neglecting two-electron term, i.e. from the parts $T_e+E_{eN}$. Phonon term comes from a second quantization of the B-O equation (7), i.e. $T_N+E_{NN}+E_e$. The entire field formulation of equation (1) is based on a hierarchical quantization of one and the same Schrödinger equation, where the electron states are quantized first, and finally nuclear states in the effective adiabatic electron potential. And these nuclear states produce vibrational states, where we solve the system of coupled oscillators with the kinetic energy $T_N$ and potential energy $E_{NN}+E_e$.

This greatly simplified case of the derivation of a field from mechanics in the B-O approximation led to the satisfactory conviction that the description of the system of electrons and nuclei is clearly determined in quantum mechanics with a single Schrödinger equation, and that field equations of the type (1) do not constitute any new ontological knowledge, but are only new mathematical expressions of the same nature as those derived directly from quantum mechanics, and that they are only an application of field methods borrowed from quantum electrodynamics, nothing more. We know that quantum electrodynamics is not derivable from quantum mechanics and vice versa. Therefore we ask: Is the relationship of mechanics a field in electron-nuclear systems so simple as it is universally believed?

## 3. Critique of the B-O separation procedure

In today's computer equipment it is no longer a problem to calculate the simplest molecules with absolute precision, without the B-O separation procedure, i.e. without the mutual separation of electrons and nuclei. Monkhorst [10,11] designed for this purpose the concept of the molecular atom. Electrons and nuclei are considered on exactly the same footing: Electrons, as well as the nuclei are counted as fermions with exchange interactions arising from the principle of indistinguishability. This is analogous to the core with a heavy particle in the center of the inertial system. It is interesting that from a philosophical point of view this concept completely bypasses the classical image of hierarchy of elementary particles → atoms → molecules. Atoms do not exist in this conception at all. The molecules themselves are "atomic molecules" or "molecular atoms", composed directly from elementary particles - electrons and nuclei.

Although this very interesting concept found few followers [12], for most scientists has remained unnoticed and ignored, for two reasons. First, it is impossible directly derive analytic expressions for measurable quantities, as is the case with B-O separation procedure, and second, even with the best computer equipment the calculations are numerically very demanding so that in practice one cannot go above ten considered particles, electrons and nuclei included. So unfortunately it is not yet possible to calculate electronically degenerated systems, such as J-T effect, where this method would surely lead to very interesting results.

On the other hand, the pragmatic point of view of a small applicability cannot be a reason for underestimation of the Monkhorst's concept. That is the only ontologically correct concept of quantum-mechanical solution of a system of electrons and nuclei, where all particles are equal regardless of whether one is heavier or lighter. It guarantees the correct solution from the viewpoint of the centre-of-mass (COM) problem; it is a warrant of the exact solutions without the necessity of the separation of particles based on differences in the values of their masses.

Most people have an opinion of the B-O approximation as the approximate solution of the Schrödinger equation, which gives very good results for 90% of all molecular and crystallic systems. In classical physics we are familiar with the approximate solutions for large systems very well. We know that we understand the laws under which these systems work, but we cannot apply them with an absolute precision, so for specific calculations we resort to the approximate solutions. On the contrary the quantum world is very treacherous. There may very easily happen that the approximate solution as a jack-o'-lantern leads us into the swamp, because behind this approximate solution we are not able to see a possible new law of nature.

The most precise analysis of the B-O approximation was made by Kutzelnigg and in his work [13] he has concluded that this was indeed an approximation designed for molecules, but in its essence is more suitable for atoms. Atoms can be solved in two ways, one is accurate, the second approximate. The exact one lies in the introduction of the atomic centre of mass, and of the relative coordinates and masses of nuclei and electrons. Transferring this exact method to the level of molecules just corresponds to the Monkhorst's concept. The approximate one is based on the fact that the mass ratio of electrons and nuclei $m/M$ is very small, so we can afford to fix the atomic nuclei and to solve electron motions in the laboratory coordinates.

At first sight it would seem that nothing precludes the philosophy, on which the B-O separation procedure is based, could be applied in the world of molecules and crystals. It would be so if the solutions of all these systems were relied solely on the above mentioned ratio *m/M*. The real problem, however, becomes the ratio of $\omega/\Delta\varepsilon$, which determines the validity of the B-O approximation, i.e. the ratio, which has no analogy in the world of atoms, because the vibrational modes do not exist there as in molecules and crystals.

The position of the B-O approximation in the scientific world is really curious. On the one hand it is something worse than an approximation, because it describes only a certain range of adiabatic systems, and there is not fully clear what holds for non-adiabatic systems for which there is no other alternative, no other well defined adequate approximation. On the other hand, it's better than an approximation, since for decades after its re-examination no one was successful to find a better one, and it's even universal in the sense that, if it's valid, so it's valid for all systems as well, and there is no need to create separately specific approximations for each system.

If so, then it follows from here that the B-O approximation is a limiting case of some higher theory, which generally describes all systems, as well as Newton mechanics is no approximation, but a limiting case of relativistic concept that includes all systems. It is therefore necessary to reach a similar higher concept that would give the same results in the adiabatic case as the B-O approximation. Such adiabatic solution will be no longer an approximation, but the limiting case of the above concept in a similar relationship, such as Einstein and Newton mechanics.

One important advantage the B-O approximation still has. It is able to reflect the hierarchy of elementary particles → atoms → molecules / crystals in the sense that corresponds to our pragmatic experience. The exact COM quantum-mechanical solution does not have this property; there the concepts of atoms and molecules coalesce into a single concept of "molecular atoms" or "atomic molecules". Now we can ask: So what's actually a real meaning of the hierarchy of elementary particles - atoms - molecules? Should we sacrifice this hierarchy in favour of exact COM quantum-mechanical solution, or look for another interpretation of this hierarchy as defined in a completely different logic than the classic one?

**4. Centre-of-mass problem**

We will try to derive a general field Hamiltonian at the ab-initio level. This means that it must be formulated and solved exclusively under the field pattern. The field Hamiltonian in the B-O approximation (1) mixes together two patterns: mechanical and field ones. From the original electron-nuclear Hamiltonian (2) we reach the B-O approximation under the mechanical pattern. Then we take over field methods from the quantum field theory, where the electron-photon interaction is replaced by the electron-phonon one. Finally the newly acquired field can be solved under the field pattern. However, mixing these two patterns is not correct. Therefore, let us start with the analysis of a breaking point, where after reaching the B-O approximation the mechanics ends and the field begins.

The last step performed under the mechanical pattern is the calculation of vibrational frequencies based on the equation (6). At this level the known COM problem arises, which is completely resolved only for atoms, but in molecules using the B-O approximation there is a lot of problems [13], and in solids it is totally ignored.

Whenever a COM problem in quantum mechanics occurs, it is always solved by the classical model. For the considered objects relative coordinates and masses are introduced, resulting in the translational or rotational movements of the entire system with reference to the center of gravity, and in the properties of the system, following from internal degrees of freedom. The quantum mechanics of molecules uses the same Hamilton-Jacobi formalism for the equations of motion as in the case of atoms. Equation (6) leads to the solution of coupled oscillators, where relative coordinates represent normal coordinates of vibrational modes. After the introduction of normal coordinates $B_r = b_r + b_r^+$ and $\tilde{B}_r = b_r - b_r^+$ for the kinetic and potential energies of nuclei in the effective field of the electrons we have

$$H_{BO} = E_{kin}(\tilde{B}) + E_{pot}(B) \tag{8}$$

where for the kinetic and potential energies it holds

$$E_{pot} = \frac{1}{4}\sum_{r \in V} \hbar\omega_r B_r^+ B_r \tag{9}$$

$$E_{kin} = \frac{1}{4}\sum_{r \in V} \hbar\omega_r \tilde{B}_r^+ \tilde{B}_r \tag{10}$$

From the B-O separation we finally get the well-known vibrational Hamiltonian

$$H_{BO} = \frac{1}{4}\sum_{r \in V} \hbar\omega_r \left(B_r^+ B_r + \tilde{B}_r^+ \tilde{B}_r\right) = \sum_{r \in V} \hbar\omega_r \left(b_r^+ b_r + \tfrac{1}{2}\right) \tag{11}$$

If instead of a system of electrons and nuclei in the B-O separation we had a simplified system of particles that are mutually coupled by means of harmonic potentials, everything would be fine. There would be a clear separation of external and internal degrees of freedom, where the internal degrees of freedom would be represented by equation (8), and, regarding the external degrees of freedom, the translations would correspond to de Broglie wave of the COM and rotations to quantized states of angular momentum with eigenvalues $L^2$ and $L_3$.

The situation is more complicated when we have not the above simplified model of particle-bound harmonic potential, but a real system consisting of electrons and nuclei. It is clear that due to the reverse influence of the electrons upon nuclei the equation (8) does not represent the exact determination of the system COM. It is generally known fact. And every one asks: Where exactly is the COM located? But no one would raise the question: What is actually the COM in the quantum world?

If we formulate the problem as a system of interacting electrons and nuclei, we cannot thereout deduce the interaction of electrons with the properties of the system, such as vibrations. The mutual interaction is allowed only for entities of the same nature, and the entities can never interact with the properties of themselves. It is a metaphysical nonsense. If we want to bring electrons and vibrations into the interaction, we have to assign an independent ontological status to vibrations, and it is not possible in dealing with a single Schrödinger equation, but via the consistent application of Bohr's complementarity principle [1,2] for the transition from mechanical to field formulation of the problem.

Instead of nuclear movements as a source of vibrations we choose atomic centers, which are not objects, but properties. Admittedly we get the equation for vibrations, which has the exact same form as that derived from the Schrödinger equation in the B-O separation, i.e. the equation (8). But it's heaven and hell, its nature as well as the solution are completely different from (8). It is the inverse equation in the sense that we do not count the properties of objects, but we create objects - vibrations from properties. So it is a standalone creative (or generic) equation, completely independent on the original Schrödinger equation for a system of electrons and nuclei, and is not derived from it, as contrasted to B-O separation.

As for solution, we encounter the terrific surprise. If the vibrations are objects, the COM becomes an object, too, and is no longer property of the electron-nuclear system, as in mechanical pattern. As a consequence, the potential energy is determined in the same way as the equation (9), i.e.

$$E_{pot} = \frac{1}{4}\sum_{r \in V} \hbar\omega_r B_r^+ B_r \qquad (12)$$

but instead of the kinetic energy (10) we get the new equation [20]

$$E_{kin} = \frac{1}{2}\left(\frac{1}{2}\sum_{r \in V} \hbar\omega_r + \sum_{r \in R}\rho_r + \sum_{r \in T}\tau_r\right)\tilde{B}_r^+ \tilde{B}_r \qquad (13)$$

where except the vibrational degrees of freedom moreover the quanta of rotational and translational degrees of freedom appear that are normally obtained from the solution of the equation for the system of coupled oscillators, but that have absolutely no meaning in the B-O separation. It simply omits these terms, because if we understand COM as a property, then such a COM leads to de Broglie wave for translations and to quantization of $L^2$, $L_3$ angular momenta for rotations. But here we are dealing with a "materialization" of COM, when as the object turns into five or six particles, which we can call rotons and translons. There is no reduction of the system to a subsystem with *3N-5(6)* degrees of freedom here, as in the B-O separation, but we must consider all *3N* degrees, and instead of vibrations we introduce a concept of hypervibrations (vibrations + rotations + translations) and the corresponding hypervibrational double-vector

$$\boldsymbol{\omega} = \begin{pmatrix}\omega_r \\ \tilde{\omega}_r\end{pmatrix} = \begin{pmatrix}\omega_r & 0 & 0 \\ \omega_r & \frac{2}{\hbar}\rho_r & \frac{2}{\hbar}\tau_r\end{pmatrix} \qquad (14)$$

whence we get the covariant expression for the boson hypervibrational Hamiltonian with respect to all *3N* hypervibrational modes.

$$H_B = \frac{1}{4}\sum_r \left(\hbar\omega_r B_r^+ B_r + \hbar\tilde{\omega}_r \tilde{B}_r^+ \tilde{B}_r\right) \quad (15)$$

Let us notice how the equation (15) differs from the analogous B-O equation (11). Although both equations are of the same type they have completely different solutions due to different interpretations of the COM in mechanics and field. The field equation (15) is not directly solvable as the mechanical equation (11), which was created as a final product of the B-O separation, but it's a gateway from the mechanics into the field, and only after the implementation in the final field equations its exceptional significance will be revealed.

## 5. General field Hamiltonian

Now we show how to implement the two equations - the mechanical equation for the system of electrons and nuclei (2) and the independent equation for hypervibrational quanta, originating from the field quantization of motion of the atomic centers (15), into one single field equation. We will not therefore solve the equation (2) according to the B-O separation procedure, but we will solve this final field equation. This means the solution of the system neither under the mechanical pattern nor under some mix of them, but solely under the field pattern. We show that the solution under the field pattern is completely different reality of the same system, studied under the mechanical pattern. This is a new type of Bohr's complementarity, arising in many-body systems. It's similar to the single-particle case where localized particles measured under mechanical pattern represent a different - complementary reality to the waves measured under field pattern.

We start with the equation (2) in the second quantization formalism. We get

$$H = T_N(\tilde{B}) + E_{NN}(B) + \sum_{PQ} h_{PQ}(B) a_P^+ a_Q + \frac{1}{2}\sum_{PQRS} v_{PQRS}^0 a_P^+ a_Q^+ a_S a_R \quad (16)$$

where one-electron matrix elements $h_{PQ}$ comprise electron kinetic energy and electron-nuclear potential energy, i.e. the terms $T_e + E_{eN}$ in the equation (2). To the equation (16) we affix the equation (15), which is the gateway from quantum mechanics to quantum field. The resulting field equation should contain solely electron and hypervibration terms, so we have to get rid of nuclear terms in the equation (16). Therefore we rewrite the equations for the potential and kinetic energies of hypervibrations as follows: In the potential energy we include internuclear potential $E_{NN}$ and some unknown potential $V_N$, arising from the electron-nuclear and electron-electron potential energies.

$$E_{pot} = E_{NN}(B) + V_N(B) \quad (17)$$

To the kinetic energy we include kinetic energy of nuclei $T_N$ and some unknown kinetic energy $W_N$ coming from the reverse influence of the electron motions.

$$E_{kin} = T_N(\tilde{B}) + W_N(\tilde{B}) \tag{18}$$

Substituting expressions from the equations (17,18) and (12,13) into the equation (16) for $T_N$ and $E_{NN}$, and using the equation (15) we get finally [20]

$$H = \sum_{PQ} h_{PQ}(B) a_P^+ a_Q + \frac{1}{2} \sum_{PQRS} v_{PQRS}^0 a_P^+ a_Q^+ a_S a_R$$

$$-V_N(B) - W_N(\tilde{B}) + \frac{1}{4} \sum_r \left( \hbar \omega_r B_r^+ B_r + \hbar \tilde{\omega}_r \tilde{B}_r^+ \tilde{B}_r \right) \tag{19}$$

The equation (19) is a field equation for the unknown electron energies $\varepsilon_P$, hypervibrational frequencies $\omega_r$ and $\tilde{\omega}_r$, and also for the unknown potential energy $V_N$ and kinetic energy $W_N$. It is the equation for the separation of electronic and hypervibrational states, which can be solved only under the field pattern. We get quite a different interpretation of the hierarchy elementary particles → atoms → molecules / crystals, than it follows from the classical philosophy. The transition from elementary particles to atoms is determined by a mechanical pattern, while the transition from atoms to molecules and crystals is given by the field pattern. In this hierarchy we cannot apply the transitive law and say that the molecules and crystals are composed of primary elementary particles – electrons and nuclei, but of the new system of elementary particles – electrons and hypervibrations. On the other hand, the system of electrons and nuclei never leads to any hierarchy, since we get only "molecular atoms" or "atomic molecules" in the Monkhorst's concept.

It should be emphasized that this analysis is the first real explanation of the mysterious relationship between mechanics and field on the quantum many-body level. In classical physics mechanics and field are different disciplines, also in the quantum world - quantum mechanics and quantum field theory. Condensed matter physics built on quantum mechanics used so far only field methods, but now it turns out that these field methods are themselves incomplete and that the quantum field for molecules and crystals must be rigorously formulated. It results in a new type of complementarity between the mechanics and the field, allowing both patterns to place under one roof, and hence to refer to a unique quantum mechanics running either under the mechanical pattern, as was rigorously formulated by Monkhorst, or under the field pattern, as shown in this work.

## 6. Solution of field equations

Fröhlich was the first who suggested solving the field equations in an attempt to resolve the superconductivity problem in his well-known transformation [22]

$$H' = e^{-S(Q,P)} H e^{S(Q,P)} \tag{20}$$

Albeit the BCS theory [23] of superconductivity was built just on this transformed Hamiltonian, no one ever realized that dependence on the momentum operator $P$ interferes in the non-adiabatic area. If we decompose (20) into a product

$$H' = e^{-S_2(P)}e^{-S_1(Q)}He^{S_1(Q)}e^{S_2(P)} \tag{21}$$

we see here that the inner part containing $Q$ is based on adiabatic transformation, while the outer part dependent on $P$ represents non-adiabatic one. Up to now, most scientists believe that superconductivity can be explained on the adiabatic level, but how to explain that the BCS theory is based on non-adiabatic Fröhlich Hamiltonian? Fröhlich had brilliant intuition, but was unlucky that applied his transformation to the wrong Hamiltonian (1), which is used in solids and which was derived from the B-O separation. Firstly, the right step consists in use of the proper field Hamiltonian (19). Secondly, in addition to vibrational modes, Fröhlich transformation must involve also rotational and translational ones. And thirdly, Fröhlich transformation must be applied without the use of translational symmetry, because, as we shall see, until the last moment we do not know what symmetry the system under consideration will actually have. Details of demanding derivations are shown at full length in the work [20], so here we will only evaluate the results, which gives the application of the Fröhlich transformation to Hamiltonian (19).

In the works [17,18] the quasiparticle analogy of the Fröhlich transformation was used that is more transparent. Unfortunately, these transformations were originally applied to the wrong electron-vibrational Hamiltonian. It was therefore the same mistake the Fröhlich did in his first attempt to explain superconductivity. These field methods started to be developed since 1983 in good faith that solid state physicists have a well-built quantum field, and that the same electron-phonon methods can be transferred into quantum chemistry. As late as in 1998 a detailed numerical analysis performed on the molecules $H_2$, HD and $D_2$ [19] showed that the correct Hamiltonian must be electron-rotation-translation-vibrational. However, the form of these quasiparticle transformations remains valid; it must be extended only by rotational and translational degrees of freedom.

The subsequent application of two quasiparticle transformations, the first Q-dependent

$$\bar{a}_P = \sum_Q c_{PQ}(B) a_Q \qquad \bar{b}_r = b_r + \sum_{PQ} d_{rPQ}(B) a_P^+ a_Q \tag{22}$$

with the unitary conditions

$$\sum_R c_{PR} c_{QR}^+ = \delta_{PQ} \qquad d_{rPQ} = \sum_R c_{RP}^+ [b_r, c_{RQ}] \tag{23}$$

and the second P-dependent

$$\bar{a}_P = \sum_Q \tilde{c}_{PQ}(\tilde{B}) a_Q \qquad \bar{b}_r = b_r + \sum_{PQ} \tilde{d}_{rPQ}(\tilde{B}) a_P^+ a_Q \tag{24}$$

with the unitary conditions

$$\sum_R \tilde{c}_{PR} \tilde{c}_{QR}^+ = \delta_{PQ} \qquad \tilde{d}_{rPQ} = \sum_R \tilde{c}_{RP}^+ [b_r, \tilde{c}_{RQ}] \tag{25}$$

in accordance with (21) leads to a new system of fermions and bosons. The diagonalization procedures allow us to choose the optimal system where we achieve a real separation into the individual fermions and bosons with the minimal interaction between them. For one-fermion energies we get nothing new, but well-known Hartree-Fock equations, i.e. the same result as we get under the mechanical pattern.

$$f_{PQ}^0 = h_{PQ}^0 + \sum_I (v_{PIQI}^0 - v_{PIIQ}^0) = \varepsilon_P^0 \delta_{PQ} \qquad (26)$$

where the terms $h_{PQ}^0$ stem from the equation (16) after the Taylor expansion around the equilibrium position:

$$h_{PQ}(B) = h_{PQ}^0 + \sum_{n=1}^{\infty} u_{PQ}^{(n)}(B) \qquad (27)$$

For one-boson energies, representing hypervibrations, we will see a surprise. The field equations yield absolutely different results than mechanical equations in the B-O separation. In the B-O approximation there are well known Pople's equations [24] for vibrational energies, derived from the variational principle. If we write Pople's equations in the native form, we obtain for the coefficients $c_{PQ}$ of the adiabatic expansion of one-electron functions in relation to the equilibrium position

$$\varphi_P(\vec{R}) = \sum_Q c_{PQ}(\vec{R}) \varphi_Q(R_0) \qquad (28)$$

the following equation

$$u_{PQ}^r + (\varepsilon_P^0 - \varepsilon_Q^0)c_{PQ}^r + \sum_{AI}[(v_{PIQA}^0 - v_{PIAQ}^0)c_{AI}^r - (v_{PAQI}^0 - v_{PAIQ}^0)c_{IA}^r] = \varepsilon_P^r \delta_{PQ}; \ c_{PP}^r = 0 \qquad (29)$$

and for the matrix of the potential energy of nuclei

$$V_N^{rs} = \sum_I u_{II}^{rs} + \sum_{AI}\left(u_{IA}^r c_{AI}^s + u_{IA}^s c_{AI}^r\right) \qquad (30)$$

Native equations are related to the original Pople's ones something like native Hartree-Fock equations to the Roothaan ones [25]. This means that at this moment we are only interested in the general derivation, not a practical form of the equations with dependence on a movable basis of atomic orbitals, necessary for numerical calculations. In addition, for comparison with field results we use normal coordinates $r$, $s$ instead of the original real $i\alpha, j\beta$. Converting from one to another in the case of the B-O approximation is trivial.

And now for comparison the same equations that are the product of the Hamiltonian diagonalization after the application of quasiparticle transformations, as derived in [20]. We get instead of one equation (29) two coupled equations for adiabatic coefficients $c_{PQ}^r$ and non-adiabatic coefficients $\tilde{c}_{PQ}^r$

$$u_{PQ}^r + (\varepsilon_P^0 - \varepsilon_Q^0)c_{PQ}^r + \sum_{AI}[(v_{PIQA}^0 - v_{PIAQ}^0)c_{AI}^r - (v_{PAQI}^0 - v_{PAIQ}^0)c_{IA}^r] - \hbar\omega_r \tilde{c}_{PQ}^r = \varepsilon_P^r \delta_{PQ}; \ c_{PP}^r = 0 \qquad (31)$$

$$(\varepsilon_P^0 - \varepsilon_Q^0)\tilde{c}_{PQ}^r + \sum_{AI}[(v_{PIQA}^0 - v_{PIAQ}^0)\tilde{c}_{AI}^r - (v_{PAQI}^0 - v_{PAIQ}^0)\tilde{c}_{IA}^r] - \hbar\tilde{\omega}_r c_{PQ}^r = \tilde{\varepsilon}_P^r \delta_{PQ}; \ \tilde{c}_{PP}^r = 0 \qquad (32)$$

Now, notice a significant difference between the mechanical equations (29) and field equations (31,32). The mechanical equations are obtained as a limiting B-O case with the non-adibatic coefficients $\tilde{c}_{PQ}^r$ equal zero and the reduction of hypervibrations to ordinary vibrations. The mechanical equations are restricted only to the adiabatic case, and when the B-O approximation expires, we get meaningless divergent results if a degeneration /

quasidegeneration of one-electron energies $\varepsilon_P^0$, $\varepsilon_Q^0$ occurs. In contrast, field equations are valid throughout the whole range of the ratio $\omega/\Delta\varepsilon$. Therefore the field separation of the system according to electron and hypervibration states seems to be natural, rather than the mechanical separation of electron and nuclear states.

In analogy with the potential energy of nuclei (30) we get two field equations for potential and kinetic energies of the atomic centers

$$V_N^{rs} = \sum_I u_{II}^{rs} + \sum_{AI}[(u_{IA}^r + \hbar\omega_r \tilde{c}_{IA}^r)c_{AI}^s + (u_{IA}^s + \hbar\omega_s \tilde{c}_{IA}^s)c_{AI}^r] \tag{33}$$

$$W_N^{rs} = 2\hbar\tilde{\omega}_r \sum_{AI} c_{AI}^r \tilde{c}_{IA}^s \tag{34}$$

And again, we can see a striking difference. Mechanical separation of electrons and nuclei does not recognize additional kinetic energy (34), which occurs only in the field formulation, and which reflects the retroactive effect of the kinetic energy of electrons on nuclei in the case of break-down of the B-O approximation.

Unfortunately, the current approach to the J-T effect is based on attempts to cope with the separation of electron and nuclear motions [7,8], thus solving the mechanical model, which leads in the case of electron degenerations to metaphysical intersection of potential curves, where besides the vibrations according to the Pople's equations (29,30) are not defined. Existing solutions via superposition of intersecting states can never replace the non-adiabaticity reflected in coefficients $\tilde{c}_{PQ}^r$ and the reverse kinetic influence of electrons in the form of additional kinetic energy $W_N$ of the vibrating system that also appears there next to the kinetic energy of nuclei $T_N$.

## 7. Field equation for the ground state energy

From the Hamiltonian (19), using the transformation (22,24), we can also derive the correction for the ground state energy. Details are described in the previous work [20], here we will present only the result

$$\Delta E_0 = \sum_{AIr}\left(\hbar\tilde{\omega}_r |c_{AI}^r|^2 - \hbar\omega_r |\tilde{c}_{AI}^r|^2\right) \tag{35}$$

The mechanical analogy of this expression unfortunately does not exist. What exists is only the Born-Handy ansatz, which appears in equation (7), but it is valid only on the adiabatic level. To make a comparison, in the equation (35) we put all non-adiabatic coefficients $\tilde{c}_{AI}^r$ equal to 0, and so we get the adiabatic expression

$$\Delta E_{0(ad)} = \sum_{AIr}\hbar\tilde{\omega}_r |c_{AI}^r|^2 = 2\sum_{AI}\left(\sum_{r\in V}\frac{1}{2}\hbar\omega_r + \sum_{r\in R}\rho_r + \sum_{r\in T}\tau_r\right)|c_{AI}^r|^2 \tag{36}$$

which we can directly compare with the Born-Handy ansatz. In the works [19,20] the exact derivation is shown, leading to the identity of the field and mechanical equations on the adiabatic level

$$\Delta E_{0(ad)} = \langle \psi(\vec{R}) | T_N | \psi(\vec{R}) \rangle_{R_0} = 2 \sum_{AI} \left( \sum_{r \in V} \frac{1}{2} \hbar \omega_r + \sum_{r \in R} \rho_r + \sum_{r \in T} \tau_r \right) |c_{AI}^r|^2 \qquad (37)$$

Numerical verification was performed on the molecules $H_2$, HD and $D_2$. It was a big surprise that the vibrational contribution makes only 20%, and the remaining 80% consists of rotational and translational contributions, even if we consider the molecules at rest, i.e. they neither rotate nor move. This comparison is very important because the field equations used in the theory of solids and derived from the B-O approximation, keep only the first term in the equation (37), and are in stark contrast with the Born-Handy ansatz. So we have a choice: either the Born-Handy ansatz is bad, or the entire BCS theory of superconductivity, based on the Fröhlich transformation of the Hamiltonian (1) is bad. Born-Handy ansatz has been tested a thousand times, and gives completely accurate results in agreement with experiment. The motive why Born-Handy ansatz has been so carefully verified was the COM problem in the B-O separation, and Handy's merit consisted in the convincing the wide scientific community of this pragmatic ansatz, without having to solve the COM problem demanding the introduction of relative coordinates and masses. Kutzelnigg then proved that the Born-Handy ansatz fully replaces the very complicated COM solution [13].

Thus we see that on the adiabatic level we get the same results from both the mechanical B-O separation and the field transformations based on hypervibrations. Hartree-Fock equations for the electron states are the same in mechanics and field as well. The field also gives us exactly the same equations for the vibrational states, which were derived by Pople in mechanics [24]. The expressions for the correction to the ground state energy are analytically different: in the mechanics it is the Born-Handy ansatz, in the field then the hypervibrational expression (36) with a somewhat more demanding evidence leading to the identity of these terms according to the equation (37). This equation (37) is therefore of paramount importance as a bridge between the mechanics and the field. The gateway from mechanics into the field is determined by the equation (15), and on the adiabatic level mechanics and field run in a parallel way, and there is a bridge (37) between them. But once the adiabatic approximation ceases to be valid, the whole mechanics based on the B-O separation ceases to be valid as well, and only the field hypervibrational equations remain valid. The mechanics is then meaningful only without the B-O separation, i.e. in the sense of the Monkhorst's concept. But on the non-adiabatic level no more bridge exists between the mechanical Monkhorst's concept and the field hypervibrational one, and both concepts lead to different results. This means that the same system studied under either mechanical or field patterns gives us a different description with different results. This is a new type of Bohr's complementarity, which is well hidden deep inside the composite systems, so it could not be discovered at the beginning of the development of quantum mechanics, unlike the evident complementarity between the particle and wave at the single-particle level.

Since the field equation (1) used in the solid state physic was derived from the B-O separation, it contains only vibrations but not hypervibrations. Fröhlich, therefore, did not derived for the ground state energy the equation (35), as it should be the correct result, but only a truncated portion

$$\Delta E_0 = \sum_{AI, r \in V} \left( \hbar \omega_r | c_{AI}^r |^2 - \hbar \omega_r | \tilde{c}_{AI}^r |^2 \right) = \sum_{AI, r \in V} | u_{AI}^r |^2 \frac{\hbar \omega_r}{(\varepsilon_A^0 - \varepsilon_I^0)^2 - (\hbar \omega_r)^2} \qquad (38)$$

from which, after substitution from quantum chemical to solid state physics notation, we get exactly the result, as it was first time derived in the Fröhlich's works [21,22].

$$\Delta E_0 = 2 \sum_{\mathbf{k}, \mathbf{k}'; \mathbf{k} \neq \mathbf{k}'} | u^{\mathbf{k}'-\mathbf{k}} |^2 f_\mathbf{k} (1 - f_{\mathbf{k}'}) \frac{\hbar \omega_{\mathbf{k}'-\mathbf{k}}}{(\varepsilon_{\mathbf{k}'}^0 - \varepsilon_\mathbf{k}^0)^2 - (\hbar \omega_{\mathbf{k}'-\mathbf{k}})^2} \qquad (39)$$

Unfortunately, this equation did not offer the expected result for the superconducting gap, as Fröhlich initially expected after the optimalization of occupation factors $f_\mathbf{k}$ when he got some decrease of the total energy and tried to interpret this new state as the state of superconductors. But if we start from the equation (35) we get after neglecting of two-electron interactions from the equations (31,32) explicit expressions for the coefficients $c_{PQ}^r$ and $\tilde{c}_{PQ}^r$

$$c_{PQ}^r = u_{PQ}^r \frac{\varepsilon_P^0 - \varepsilon_Q^0}{(\hbar \omega_r)^2 - (\varepsilon_P^0 - \varepsilon_Q^0)^2} \qquad (40)$$

$$\tilde{c}_{PQ}^r = u_{PQ}^r \frac{\hbar \tilde{\omega}_r}{(\hbar \omega_r)^2 - (\varepsilon_P^0 - \varepsilon_Q^0)^2} \qquad (41)$$

and herefrom the correct expression for the ground state energy correction

$$\Delta E_0 = \sum_{AIr} | u_{AI}^r |^2 \frac{\hbar \tilde{\omega}_r}{(\varepsilon_A^0 - \varepsilon_I^0)^2 - (\hbar \omega_r)^2} \qquad (42)$$

that in the form of the sum of vibrational, rotational and translational parts finally reads

$$\Delta E_0 = \sum_{AI, r \in V} | u_{AI}^r |^2 \frac{\hbar \omega_r}{(\varepsilon_A^0 - \varepsilon_I^0)^2 - (\hbar \omega_r)^2}$$
$$+ 2 \sum_{AI, r \in R} | u_{AI}^r |^2 \frac{\rho_r}{(\varepsilon_A^0 - \varepsilon_I^0)^2} + 2 \sum_{AI, r \in T} | u_{AI}^r |^2 \frac{\tau_r}{(\varepsilon_A^0 - \varepsilon_I^0)^2} \qquad (43)$$

Detailed discussions how to solve the equation (43) were described in the work [20]. Here we mention the most important results concerning the degeneration and structure of systems. From the equation (43) it can be seen that just matrix elements of electron-roton and electron-translon interaction, whether or not they are all zero, determine, whether or not the system remains degenerate / quasidegenerate. Zero coefficients lead to a metallic phase and the system behaves adiabatically. The hypervibrational equation (43) reduces to a simple vibrational one (38), which in quasimomentum notation has exactly the same form (39) as derived first by Fröhlich. This equation can never explain the ground state of superconductors, as initially Fröhlich supposed; it is only the equation for the correlation energy of conductors.

On the other hand, the non-zero coefficients lead to a superconducting ground state, which is non-adiabatic. It is extremely important to the last step to write all the equations without the introduction of symmetry, i.e. in the general quantum-chemical notation, because the systems under the field pattern may have a different symmetry than under the mechanical pattern, which is evident in superconductors, including low-temperature ones. Low-temperature superconductors are generally considered to be adiabatic, but it's not true. Symmetry can be deceptive. Under the mechanical pattern the same internuclear distances of the material are measured that is conductive above the critical temperature and superconductive below the critical temperature, and consequently it is considered that the symmetry after the transition to the superconducting state is maintained. But under the field pattern we find the opposite: Roton and translon terms in (43) cause a slight change of internuclear distances due to a singularity in the original symmetrical configuration, and the symmetry of the crystal after the transition to the superconducting state is reduced by half. So the bands are no longer half-filled as in conductive state, but fully occupied as in insulators.

We denote the unperturbed energies of the lower valence band as $\varepsilon_{v,\mathbf{k}}^0$, and those of the higher conducting band as $\varepsilon_{c,\mathbf{k}}^0$. In a similar way we get twice as many hyperphonon branches – innerband with accoustical branches, and interband containing only the optical branches, but moreover rotons and translons. We denote the frequencies of the former set as $\omega_{a,\mathbf{q}}$ and the frequencies of the latter set as $\omega_{o,\mathbf{q}}$. Finally the vibronic coupling via the optical phonon modes stabilizes the whole system in this new configuration. After the rewriting of the equation (43) in solid state notation ($r \to o,\mathbf{q}$; $I \to v,\mathbf{k},\sigma$; $A \to c,\mathbf{k'},\sigma'$; $\varepsilon_I^0 \to \varepsilon_{v,\mathbf{k}}^0$; $\varepsilon_A^0 \to \varepsilon_{c,\mathbf{k'}}^0$; $u_{AI}^r \to u_{\mathbf{k'k}}^{\mathbf{q}} = u^{\mathbf{k'-k}} = u^{\mathbf{q}}$) we get

$$\Delta E_0 = 2 \sum_{\mathbf{k},\mathbf{k'}} |u^{\mathbf{k'-k}}|^2 \frac{\hbar \omega_{o,\mathbf{k'-k}}}{(\varepsilon_{c,\mathbf{k'}}^0 - \varepsilon_{v,\mathbf{k}}^0)^2 - (\hbar \omega_{o,\mathbf{k'-k}})^2}$$

$$+ 4 \sum_{\mathbf{k},r \in R} |u^r|^2 \frac{\rho_r}{(\varepsilon_{c,\mathbf{k}}^0 - \varepsilon_{v,\mathbf{k}}^0)^2} + 4 \sum_{\mathbf{k},r \in T} |u^r|^2 \frac{\tau_r}{(\varepsilon_{c,\mathbf{k}}^0 - \varepsilon_{v,\mathbf{k}}^0)^2} \qquad (44)$$

Complementarity between the mechanical and field patterns thus leads in non-adiabatic systems to different understanding of symmetry, and thus the structure. The "field" symmetry may differ from the "mechanical" symmetry, and this enigma is intrinsic just to superconductors.

## 8. Comparison with the special theory of relativity

As shown in equation (37), the Born-Handy ansatz in the CPHF form is relativistic in regard to the inseraparable set of internal and external degrees of freedom. And the same relativistic expression can be derived directly under the field pattern, without the need of knowledge of the Born-Handy ansatz and its derivation from the mechanical pattern, in exactly the same style as now Maxwell equations can be formulated relativistically using the four-dimensional potential.

Including the center of gravity movement leads to the principle of inseparability of degrees of freedom, and thus a new form of relativity, which applies to structure of condensed matter. The treatment of the field equations is nonperturbative and reaches solutions on the one-particle level.

The non-perturbativity is guaranteed due to divergence of COM terms in relativistic field equations in the degenerate cases. It results in a stable nonperturbative solution where the degeneration is completely bypassed, and there is no need to count multideterminant perturbations in order to remove the degeneracy on the principle of superposition. And it has the important consequence: the principle of relativity is superior to the principle of superposition. Does this mean that the principle of superposition is not valid? No, it holds still valid, but is suppressed.

All what we can do now, is the formal comparison of the presented theory with the special theory of relativity:

\*   The degrees of freedom - vibrations, rotations and translations - are inseparable in principle as well as space and time in the special relativity.

\*   The covariant notation for all operations applied to the degrees of freedom can be constructed in a similar way as the space-time covariant notation in the special relativity.

\*   The Born-Handy formula in quantum mechanics plays a similar role as the Maxwell equations in classical physics: the former leads to COM quasiparticle transformations, the latter to the Lorentz transformations.

\*   The crude representation is the only one where the degrees of freedom are fully separated; it corresponds to the reference system in the special relativity, only where the components of space and time are separated.

\*   All other representations after the application of COM quasiparticle transformations link all degrees of freedom in one inseparable set as well as the inertial systems after the application of Lorentz transformation link space a time.

\*   The ratio $\omega/\Delta\varepsilon$ has the same role as the ratio $v/c$ in the special relativity, and the values of both of them are restricted. Whereas the latter ratio leads to limited velocity, the former implies the bypassing of electron degeneracies.

\*   Both theories have "classical" limits. The case $\omega/\Delta\varepsilon \ll 1$ corresponds to the adiabatic limit with "classical" definition of structure based on the B-O approximation, as well as the case $v/c \ll 1$ implies the validity of the classical Newton mechanics. Then the absolute meaning of degrees of freedom corresponds to the absolute meaning of space and time.

\*   Both theories have ultrarelativistic ranges. The non-adiabatic effects in the presented theory are "ultrarelativistic" in this sense. The structure cannot be defined in a simple

"classical" way by means of the B-O approximation. Then the relative meaning of degrees of freedom corresponds to the relative meaning of space and time.

The "ultrarelativistic" case of breakdown of B-O approximation is the most interesting topic of the presented work. Since the molecular and crystallic structure is a direct consequence of the separation of coordinates into the vibrational, rotational and translational degrees of freedom, and these degrees of freedom have only relative meaning in quantum mechanics running under the field pattern, we shall call the presented theory as the theory of relativity of structure, in order to distinguish it from the Einstein's relativity.

Let us continue with the ontological paradigmatic background. The classical principle of relativity, connected with the introduction of relative coordinates within the COM separation process, cannot be lost after the replacement of the COM separation with the Born-Handy ansatz. The relative meaning of degrees of freedom is just the quantum field reflection of the classical principle of relativity. In other words, the classical principle of relativity in Newton mechanics implies the principle of inseparability of degrees of freedom in quantum mechanics running under the field pattern. How differ solid bodies in classical and quantum mechanics? The properties of a classical solid body are defined through the internal degrees of freedom, i.e. vibrations, whereas the external degrees (rotations + translations) determine the position of the body in relation to the rest of the universe. Quantum mechanics is holistic, and therefore cannot define precisely the solid body, separated from the rest of the universe. The degrees of freedom have only the relative meaning and the internal degrees are mixed with the external ones. So the small light electron can interact via the electron-roton and electron-translon interaction with the whole macroscopic crystal. It is a similar holistic effect as the EPR paradox [26]. And just here we can find the true explanation, why superconductivity is a microscopical effect on macroscopical level.

## 9. Conclusion

In this work we have shown that field methods cannot be transferred from quantum field theory into quantum mechanics where the B-O approximation is the starting point. The main reason is the fact that such field methods lead to quite erroneous results in comparison with the Born-Huang ansatz, i.e. already at the adiabatic level. Unfortunately, there are no means of reconciliation of such a field with the next corrections above the B-O approximation. These methods can therefore never properly capture non-adiabatic phenomena such as superconductivity. The author dealt many years with the possibility of extending the solid-state field methods into the quantum chemistry, but numerical tests of hydrogen and deuterium molecules turned out that these field methods differ almost in one decimal place from the real values provided by the Born-Huang ansatz. This ultimately led the author to renounce entirely field methods and work on the construction of an independent and standalone field.

1) It was therefore necessary to perform a generalized formulation of the field without using the B-O approximation. This means that for the calculation of vibrational energies we cannot use mechanical means, where the vibrations are interpreted as properties (eigenstates) of the system composed of objects (moving nuclei in the effective field of electrons). If we want to introduce a field where the electrons interact with the vibrational quanta (phonons), these vibrational quanta must have the same ontological status as electrons; hence they have to be objects, not properties acquired from the B-O approximation. We chose ansatz - the atomic centers as a source of vibrational quanta, where these quanta correspond to unknown kinetic and potential energies, obtained not from the B-O approximation, but solely from selfconsistent field equations, without using any mechanical means. This leads to the formulation of an independent and standalone field as another possible description of composite systems, in addition to the already known mechanical description in the form of fundamental equation (2). Since the electronic and vibrational spectra are well measurable, the separation of electronic and vibrational states under the field pattern is meaningful and well analytically solvable on the whole range of vibrational and electronic energies. On the other hand, the mechanical pattern must rely on the separation of electron and nuclear motions, which is due to the small ratio $m/M$ partly possible in the form of the B-O approximation, however, valid only in the case of a small ratio of vibrational and electronic energies. In the general case the separation of electron and nuclear motions is yet impossible under the mechanical pattern, thus the only correct method is the Monkhorst's concept, including the COM problem and introducing relative coordinates and masses. Separation of electron and nuclear motions in quantum mechanics is not an archetype; the separation of electronic and vibrational states in the concept of an independent and standalone field pattern becomes archetype. This leads to a new form of manifestation of the Bohr's principle of complementarity on the level of composite systems.

2) In field methods applied within the range of the B-O approximation, the COM problem was never considered because the secular equation for the vibrational frequencies was handled under the mechanical pattern, solving the COM problem according to the rules of mechanics, and the field methods were built on this mechanical outcome. But if we consider an independent and separate field with the exclusion of all mechanical treatments, we are forced to deal with the COM problem on a pure field level. While the mechanical pattern provides us with the vibrational states as a property, the field pattern works with vibrational states as objects, in solids known as phonons, which enter into the interaction with electrons. The same rule holds for the center of gravity, too. In a study of a system under the mechanical pattern, we ask: Where is the center of gravity? Under the field pattern, however, we must ask: What is the center of gravity? The center of gravity is represented by five or six quasiparticles - rotons and translons. This has an analogy neither in the classical physics, nor in the classical logic. Solving of the COM problem under the field pattern is no longer subordinated to mechanical Hamilton-Jacobi formalism, seeking the right COM position, but there is a mysterious metamorphosis - vibrations as well as COM materialize and transform into the hypervibrations or hyperphonons - quasiparticles composed of phonons, rotons and translons. This has a dramatic consequence: All treatments of the field Hamiltonian must be performed in a covariant way with respect to all 3N degrees of freedom, where the rotational and translational degrees of freedom cannot be separated and excluded. Electron-vibrational or electron-phonon Hamiltonian is therefore not an archetype; the electron-hypervibrational or electron-hyperphonon field Hamiltonian becomes archetype. This leads to a new type of relativity of internal and external degrees of freedom, having many common features with the special theory of relativity of space and time.

3) Field methods, taken from quantum field theory and used in solids, are based on the law of quasimomentum conservation. Unfortunately, only a certain subset - crystals with translational symmetry, meets the requirements for the introduction of quasimomenta for electrons and phonons. A general field has to be formulated not only for crystals, but also for all molecules exhibiting no translational symmetry. And even in crystals with translational symmetry, it is necessary to perform all derivations until the last stage without the introduction of any symmetry, as shown on the applications of the final groundstate equation (43), which has in the case of crystals with translational symmetry two solutions, one for conductors (39) and the other for superconductors (44). As late as the final form of the equations shows us what a real structural symmetry the crystal actually has. This is extremely important for a proper understanding of the symmetry breaking. The apriori given law of momentum conservation related to the translational symmetry is therefore not an archetype for a general formulation of the field in condensed matter physics, unlike quantum electrodynamics; the general field without any symmetry, where electron states are expressed via quantum-chemical spinorbitals and hypervibrational states via indexes of these modes, and where the possible symmetry is aposteriori obtained as a solution of field equations, becomes archetype. This leads to an utterly revised view of symmetry breaking in non-adiabatic systems, namely J-T effect and superconductivity.

Summarizing, we have revealed three fatal errors incurred from a blind transferring of quantum field methods into the quantum mechanics. This had tragic consequences because the model Hamiltonian (1) was considered sufficient for a description of solids including superconductors. From there, of course, Fröhlich derived wrong effective Hamiltonian, from which incorrect BCS theory arose. Thus, everything that was derived on a microscopic level regarding superconductivity is necessarily bad, except the Josephson effect, where the derivation is independent on its specific microscopic mechanism.

On the other hand, if we realize these errors, we open the way for construction and understanding of the true form of field, generally applicable to all molecules and crystals. Only atoms are described by means of a single mechanical Hamiltonian (2). But from the simplest molecule $H_2^+$ up to arbitrary complex systems moreover the field comes into play, and so molecules and crystals can be described in two ways: in addition to mechanical Hamiltonian (2) we have a general field Hamiltonian (19). This field Hamiltonian is firstly independent, which means that the hypervibrational energies are not derived from the mechanical B-O approximation, but as unknown variables bound to the atomic centers are calculated from purely selfconsistent field equations. Secondly, this field Hamiltonian is standalone, which means that it is sufficient and complete for the description of molecules and crystals. This dual expression of molecules and crystals either as a mechanical system of electrons and nuclei, or as a field of electrons and hypervibrations, leads to a new type of Bohr's complementarity.

As shown in this article, Bohr's complementarity, observed at the single-particle level, extends also to the nature of composite systems. As an elementary entity can be observed in complementary aspects once like a particle, next time like a wave, depending on if it is studied under the mechanical pattern or under the field one, the same complementarity we also meet in composite systems that can run under both mechanical as well as the field pattern. The classic rules taken from the classical mechanics, allowing from the knowledge of the equations of motion for each component simply to write motion equations for the whole system, are no longer valid. The classical determinism is broken. Under the mechanical pattern we get on the basis of the precise Monkhorst's concept only "atomic molecules" or "molecular atoms", but never a hierarchy of elementary particles → atoms → molecules. This hierarchy, and thus explanation of the existence of molecules and crystals, is possible only after the introduction of the field Hamiltonian (19).

This raises an important question: Why such an important phenomenon as Bohr's complementarity in composite systems has never been discovered. Maybe there are two reasons for this. Firstly, the orthodox interpretation of quantum mechanics, known as the Copenhagen interpretation and embracing Bohr's complementarity [1,2], had many years to fight for its survival. Even some famous physicists, such as Einstein [26], disagreed with it. Many models of so-called hidden parameters have been developed, but without any success. To this day, there are groups of scientists, pursuing its revision in favor of deterministic formulation of the microworld. Copenhagen interpretation ended up with hydrogen atom and it was assumed that for composite systems the classical logic will be sufficient as in classical mechanics, where from the knowledge of the equations of motion for single particles we can generally derive equations for the entire composite system. Secondly, the B-O approximation played the negative role in the development of our knowledge of microworld since it hides this new manifestation of the Bohr's complementarity. This approximation, despite its success and widespread use, is a merely pragmatic rule, and not the law of nature. It gives an illusion that it is able to explain the world of molecules and crystals. But it is ontologically incorrect.

It was shown that the hierarchy from atoms to molecules and crystals is only possible under the field pattern, as it was developed in previous work [20]. How in this context seems to be the B-O approximation? It tries to solve under the mechanical pattern the hierarchy from atoms to molecules, which is reserved only for the field pattern. Its success is limited solely to the adiabatic cases, where the mechanical and field patterns lead to the same results, i.e. in the case of $\Delta\varepsilon/\omega \gg \hbar$. It's similar to the single-particle case where in the classical limit given by the product $2\Delta q \Delta p \gg \hbar$, i.e. far from the border defined by the Heisenberg uncertainty relation, the mechanical and field pattern lead to the same results, too.

Let us notice another important comparison of mechanical and field patterns on one side and the B-O approximation on the other side in connection with the COM problem. Mechanical and field patterns always lead to a sacrifice. Under the mechanical pattern in the accurate Monkhorst's concept we have to sacrifice laboratory system in favor of the relative masses and coordinates. Under the field pattern the laboratory system remains in force, but we have to sacrifice absolute meaning of degrees of freedom in favor of the principle of inseparability, where the structure is relativistic with respect to these degrees of freedom. Only the B-O approximation acts as a kind of "classical limit", where we do not have to sacrifice anything. We can continue to work in a laboratory system with degrees of freedom, having an absolute meaning, that is, based on the classical ideas about the structure of molecules and crystals from the 19th century. The B-O approximation played for years the role of a frog sitting on a spring, and hid the new quantum world, emerging at the composite level above the single-particle quantum world, as we know it. The model field Hamiltonian (1), which is criticized in this work, was derived from the B-O approximation in a deterministic way, so it is a many-body analogy of methods of hidden parameters at the single-particle level.

In this work a comparison of the relative meaning of degrees of freedom with the special theory of relativity was demonstrated. We have shown a similarity between Maxwell equations, transforming according to the Lorentz transformation binding together space and time, with electron-hypervibrational field equations, which are transformed in a similar style, where internal and external degrees of freedom are tied together. From the general relativity we know that the metric of space-time is influenced by the surrounding mass of the universe. Likewise, if in condensed systems we move from the mechanical to the field description, the field equations contain no more explicit nuclear contribution. A huge mass of nuclei is then reflected in the metric of degrees of freedom; internal and external degrees can no longer be separated from each other, so in the field description we are forced to use hypervibrations instead of plain vibrations.

The most interesting is the "ultrarelativistic" region, known as the break-down of the B-O approximation. It is the case of degenerate / quasidegenerate electronic levels, as we meet them at the J-T effect and superconductivity. Today more and more scientists are convinced of the connection between these two phenomena. There are many attempts to explain superconductivity on the bases of the J-T effect [9]. Both phenomena are commonly explained using the principle of superposition: namely the J-T effect by considering the mechanical pattern with two or more intersecting adiabatic surfaces, and the BCS theory by considering the field pattern with the superposition of excited states of the Cooper pairs. Both of these phenomena are associated with symmetry breaking, but that is interpreted differently: in the case of the J-T effect it is the structural symmetry breaking, and in the case of superconductivity it is the charge superselection rule violation. But if we take into account three above discussed fatal errors afflicting the model Hamiltonian (1), from which unfortunately the Fröhlich Hamiltonian was derived, and from it the BCS theory, we finally understand that the principle of relativity is superior to the superposition principle, and that there is no degeneracy of electronic states under the field pattern since in the symmetric points singularities occur, caused by electron-roton and electron-translon interactions. Therefore the systems bypass this degeneration at the one-particle level. It's no more necessary to try to explain a state of superconductivity on the bases of the J-T effect, but we conclude that these two phenomena are synonyms and share a unique common symmetry breaking, which is of the structural nature, and no charge superselection rule violation is needful.